\documentclass[12pt]{article}
\usepackage{amsfonts}
\usepackage{amsmath}
\usepackage{amssymb}
\usepackage{graphicx}
\usepackage{color}
\usepackage[all, knot]{xy}
\usepackage{tikz}
\usepackage{array}

\usepackage{ulem}

\usepackage[utf8]{inputenc}
\usepackage{epstopdf}
\usepackage[footnotesize]{caption}
\usepackage{amsthm}
\usepackage{enumitem}
\usepackage{mathrsfs}

\def \be {\begin{equation}}
\def \ee {\end{equation}}
\def \bea {\begin{eqnarray}}
\def \eea {\end{eqnarray}}
\def \nn {\nonumber}

\def \rr {\raise.35ex\hbox{\small $\prime$}\kern-.17em{\mbox{\large $\imath$}}}

\def \dels {\partial\kern-.6em /\kern.1em}
\def \As {{A\kern-.5em / \kern.5em}}
\def \Ds {D\kern-.7em / \kern.5em}

\def \ks {k\kern-.5em /}
\def \ls {l\kern-.5em /}

\newcommand{\el}[1]{\label{#1}}
\newcommand{\er}[1]{\eqref{#1}}

\newcommand{\ci}[1]{}

\newcommand{\ke}{\rangle}

\newcommand{\lb}{\left(}
\newcommand{\rb}{\right)}

\newcommand{\lsb}{\left[}
\newcommand{\rsb}{\right]}

\newcommand{\ba}{\begin{eqnarray}}
\newcommand{\ea}{\end{eqnarray}}
\newcommand{\bal}{\begin{align}}
\newcommand{\eal}{\end{align}}
\newcommand{\bay}[1]{\left(\begin{array}{#1}}
\newcommand{\eay}{\end{array}\right)}
\newcommand{\eg}{\textit{e.g.} }
\newcommand{\ie}{\textit{i.e.}, }
\newcommand{\iv}[1]{{#1}^{-1}}
\newcommand{\st}[1]{|#1\ke}

\newcommand{\zt}[1]{\textrm{#1}}

\def\xD{{\Delta}}
\def\xe{{\epsilon}}

\def\xl{{\lambda}}

\def\CM{{\cal M}}

\def\CO{{\cal O}}

\reversemarginpar
\newcommand{\hide}[1]{}

\DeclareMathOperator{\Tr}{Tr}

\makeatletter

\newlist{axioms}{enumerate}{2}
\setlist[axioms,1]{label=\textbf{A\arabic{axiomsi}.}, ref=A\arabic{axiomsi}}
\setlist[axioms,2]{label=\textbf{A\arabic{axiomsi}\rlap{\myEnumCounter{axiomsii}}.},%
                   ref=A\arabic{axiomsi}\myEnumCounter{axiomsii},%
                   align=parleft,%
                   leftmargin=0em,%
                   itemsep=1.4ex,%
                   before={\stepcounter{axiomsi}}}

\usepackage{tikz,pgf}
\usetikzlibrary{shapes}
\usetikzlibrary{calc}
\usetikzlibrary{decorations.pathmorphing}
\usetikzlibrary{decorations.pathreplacing,shapes.misc}
\usetikzlibrary{positioning}
\usetikzlibrary{arrows}
\usetikzlibrary{decorations.markings}

\tikzset{snake it/.style={decorate,decoration={snake,segment length=1.5mm, amplitude=.3mm}}}
\tikzset{biggerarrow/.style={
    decoration={markings,mark=at position 1 with {\arrow[scale=1.5]{>}}},
    postaction={decorate},
    shorten >=0.4pt}}
\tikzset{arrow at middle/.style={decoration={
    markings,
    mark=at position 0.5 with {\arrow{>}}}}}

\begin{document}

\begin{titlepage}

\begin{center}

\hfill
\vskip .2in

\textbf{\LARGE
Analysis of the Entanglement with Centers
\vskip.5cm
}

\vskip .5in
{\large
Xing Huang$^{a,b,c,d}$ \footnote{e-mail address: xingavatar@gmail.com} and
Chen-Te Ma$^{e,f,g,h}$ \footnote{e-mail address: yefgst@gmail.com}\\
\vskip 3mm
}
{\sl
${}^a$
Institute of Modern Physics, Northwest University, Xian 710069, China.\\
${}^b$
Shaanxi Key Laboratory for Theoretical Physics Frontiers, Xian 710069, China.\\
${}^c$
NSFC-SPTP Peng Huanwu Center for Fundamental Theory, Xian 710127, China.\\
${}^d$
Department of Physics, National Taiwan Normal University, Taipei 10617, Taiwan, R.O.C..\\
${}^e$
Guangdong Provincial Key Laboratory of Nuclear Science,\\
 Institute of Quantum Matter, South China Normal University, Guangzhou 510006, China.\\
${}^f$
School of Physics and Telecommunication Engineering,\\ 
South China Normal University, Guangzhou 510006, China.\\
${}^g$
The Laboratory for Quantum Gravity and Strings,\\
 Department of Mathematics and Applied Mathematics,
University of Cape Town, Private Bag, Rondebosch 7700, South Africa.\\
${}^h$
Department of Physics and Center for Theoretical Sciences,\\ 
National Taiwan University, Taipei 10617, Taiwan, R.O.C..
}\\
\vskip 3mm
\vspace{60pt}
\end{center}
\newpage
\begin{abstract}
We begin from the quantization algebras and constraint for analyzing the choice of centers in the first-order formulation without losing generality.  
Then we calculate the entanglement entropy in the non-interacting $p$-form theory in $2p+2$ dimensional Euclidean flat background with an $S^{2p}$ entangling surface. 
The universal term of the entanglement entropy in the non-interacting $p$-form theory is determined in terms of the universal terms of the non-interacting zero-form theory. We also prove the strong subadditivity in the non-interacting theory with the non-trivial centers. Finally, we calculate the mutual information with centers in two-dimensional conformal field theory. The result shows that the mutual information is independent of the choice of centers.
\end{abstract}

\end{titlepage}

\section{Introduction}
\label{Sec:1}
\noindent 
Quantum entanglement is a physical phenomenon that the quantum state of each particle cannot be described independently. When a partial trace operation acts on a density matrix, one can obtain the reduced density matrix. Then one can obtain a useful quantity, entanglement entropy, for counting the physical degrees of freedom in a subregion.
\\

\noindent 
The entropy of black hole \cite{Bekenstein:1972tm} obtained by Bekenstein-Hawking formula gave a motivation to study entanglement entropy in quantum field theory and many-body system. 
Entanglement entropy also generally satisfies the area law just like black hole entropy.
The early observation is that the horizon can be seen as an entangling surface decomposing the degrees of freedom between the interior and exterior of the black hole. Therefore, it was suggested that quantum correction between these two regions might contribute to the black hole entropy. Hence quantum correction to the black hole entropy from matter fields was conjectured to have a state counting interpretation as the entropy of entanglement \cite{tHooft:1984kcu, Bombelli:1986rw}. 
Unfortunately, this attempt failed in two-dimensional Abelian gauge theory \cite{Kabat:1995eq}. Since the two-dimensional Abelian gauge theory is topological, the theory does not have dynamics. In other words, the entanglement entropy vanishes. The black hole entropy does not vanish in a covariant gauge. The difference between the black hole entropy and entanglement entropy is a contact term, which arises from the coordinate singularity. We brought up this example to emphasize the importance of contact term \cite{He:2014gva}. 
\\

\noindent 
A generic approach in defining the entanglement entropy is to consider operator algebras that belong to von Neumann algebra \cite{Casini:2013rba, Ma:2015xes}. This approach gives a clear understanding of the connection between a Hilbert space and entanglement. We choose the centers of operators to define the separation. The centers are operators that commute with all operators in the Hilbert space. In gauge theories, it is usually necessary to introduce a nontrivial center to define spatial entanglement. This approach also introduces the extended lattice model in the lattice Yang-Mills gauge theory \cite{Buividovich:2008kq}. Introducing the center provides a new interpretation to the bipartition of Hilbert space in gauge theory without the issue of partial trace operation. 
When centers are not identity operators, the Hilbert space forms a superposition of the tensor product decomposition. Then we can choose suitable centers (like electric fields) on the entangling surface without cutting the Wilson loop. 
\\ 

\noindent 
The use of von Neumann algebra makes it easier to understand the intrinsic physical properties of entanglement, which nevertheless remains hard to compute practically. This problem also arises in many different contexts as it is usually hard to compute a quantity in the operator formalism. Nevertheless, the computation was simplified in the Lagrangian formalism, in which the choice of centers is equivalent to the choice of boundary conditions \cite{Ma:2015xes, Buividovich:2008kq, Donnelly:2015hxa, Donnelly:2014fua, Ghosh:2015iwa, Soni:2015yga, Buividovich:2008gq, Donnelly:2011hn}.\\

\noindent 
 In quantum field theory that has non-trivial centers (the trivial center being the identity operator), we can remove operators \footnote{An operator is removed in the sense that it is no longer included in $A_V \cup A_V'$, where $A_V$ is the operator $A$ in the region $V$, and $A_V'$ is the operator $A_V'$ in the region $V'$.}, which changes the algebra associated to a region and leads to non-trivial centers. When one removes operators or choose non-trivial centers in an entangling surface, it is equivalent to making a different observation to the entangling surface. The Hilbert space forms a direct sum decomposition. The decomposition has been considered in many-body systems with continuous symmetry \cite{Goldstein:2017bua, Xavier:2018kqb, Fraenkel:2019ykl}. The choice of centers only relies on the kinematic information, algebra. Therefore, this kind of analysis can be done generically without suffering from the technical issue.
\\
 
 \noindent 
 In this paper, we want to discuss four open questions from the {\it quantization, entanglement entropy, strong subadditivity, and mutual information} by working and computing explicit examples. The first question is whether the Hilbert space in topological gauge theory and Dirac fermion theory can be decomposed by non-trivial centers. Intuitively, removing operators modifies the decomposition due to the dynamics of removed operators. Therefore, topological quantum field theory with the trivial topology should not have a direct sum decomposition. We also do a similar analysis in the fermion theory and get the same conclusion. To our knowledge, this analysis has not been done before. 
\\
 
 The second question is whether the physical degrees of freedom of the gauge fields can be determined from the physical degrees of freedom of the scalar fields through the universal terms of entanglement entropy. Since gauge fields are bosonic fields, we should expect so. Because the entanglement entropy in the Abelian gauge theory can receive contributions from the edge modes, the generalization to the $p$-form theory provides the non-trivial test of the equivalence between the edge mode or Lagrangian methods and the Hamiltonian methods \cite{Casini:2014aia}. There are two different ways to do the computation. The first way is to consider the boundary terms \cite{Ma:2015xes, Donnelly:2015hxa, Donnelly:2014fua} and the second way is to consider a regularization for the zero modes of eigenfunctions in the heat kernel \cite{Donnelly:2012st}. The two methods should be equivalent and give consistent entanglement entropy from different regularization schemes. The universal term of non-interacting $p$-form theory in $2p+2$ dimensions can be understood as the sum of contributions from the bulk and boundary terms. As in the case of Abelian vector gauge theory, the sum agrees with the known anomaly coefficients, and the bulk part could be reproduced from the heat kernel on the manifold $S \times H^{2p+1}$, where $S$ is a one-dimensional sphere and $H^{2p+1}$ is the $2p+1$ dimensional hyperbolic space, which has the negative constant curvature.
\\
 
 The third question is whether the strong subadditivity still holds even if we include the non-trivial centers. The old proof was only valid for the trivial center \cite{Araki:1970ba, Lieb:1973cp}.  The strong subadditivity is related to three regions. Therefore, we need to have common centers between the three regions. Otherwise, the proof cannot be straightforwardly done for the generic case. We found that the non-interacting theory can generically satisfy the strong subadditivity because the reduced density matrix is the direct sum of the reduced density matrices of the trivial center with a probability distribution of centers. Usually, the reduced density matrices in the decomposition depend on the probability distribution of centers. Because the non-interacting theory does not suffer from this problem, we can show the strong subadditivity generically.
\\
 
 The fourth question is whether mutual information is independent of the choice of centers. We use two-dimensional conformal field theory to provide a demonstration \cite{Ohmori:2014eia, Calabrese:2004eu}. Even for the non-gauge theory, the non-universal term of entanglement entropy can be affected by the regularization. The choice of centers affects regularization terms. Therefore, the mutual information in two-dimensional conformal field theory does not have the regularization terms.
\\
 
The outline of the paper is as follows. We  analyze the decomposition in two-dimensional gauge theory, the three-dimensional Chern-Simons gauge theory, and the fermion field theory in Sec.~\ref{Sec:2}. In Sec.~\ref{Sec:3}, we calculate the entanglement entropy of the non-interacting $p$-form theory and also discuss the strong subadditivity of non-interacting theory. We calculate the mutual information with centers in CFT$_2$ in Sec.~\ref{Sec:4}.  Finally, we conclude in Sec.~\ref{Sec:5}. We review von Neumann algebra in the context of entanglement with centers in Appendix \ref{app1}, the Lagrangian formulation for the entanglement with centers in Appendix \ref{app2} and the one-form gauge theory in Appendix \ref{app3}. The details of the bulk entanglement entropy in the Abelian $p$-form non-interacting theory in Appendix \ref{app4}. We show the relation between the sphere and the two-dimensional cone in Appendix \ref{app5}.

\section{Analysis of Decomposition}
\label{Sec:2}
We study decomposition from a few examples: two-dimensional Yang-Mills gauge theory, three-dimensional Chern-Simons theory, and fermion theory, and start the analysis from the quantization algebras and constraints in the Hilbert space. The brief review of von Neumann algebra in Appendix \ref{app1}.

\subsection{Two-Dimensional Yang-Mills Gauge Theory}
The quantization algebras and constraints in the two-dimensional Yang-Mills gauge theory is given by:
\bea
&&\lbrack A_1^a(x), F_{01}^b (y)\rbrack=i\delta^{ab}\delta(x-y),\qquad \lbrack A_1^a(x), A_1^b(y)\rbrack=0, 
\nn\\
&&\lbrack F_{01}^a(x), F_{01}^b(y)\rbrack=0, \qquad D_1F^{10, a}=0,
\eea
in which the Lie algebra indices are labeled by the indices $a$-$z$ and $D$ is the covariant derivative. The field strength is defined as $F_{\mu\nu}^a\equiv \partial_{\mu}A_{\nu}^a-\partial_{\nu}A_{\mu}^a+f^{abc}A_{\mu}^bA_{\nu}^c$, in which the spacetime indices are labeled by Latin letters, and $f^{abc}$ is the structure constant of gauge group.
\\

We first choose the gauge fixing condition $A_0^a=0$, and then we have the condition $\partial_1A^{1, a}=0$. As a result, the gauge field $A_1^a$ only depends on the time. In other words, there is no entanglement in two-dimensional gauge theory unless the topology is non-trivial. 
\\

In the case of the non-trivial topology of two-dimensional Abelian gauge theory, we consider a spatial circle. We have a non-local degree of freedom, the Wilson line around the circle. The electric field in the two-dimensional Abelian gauge theory is gauge invariant for each point of the circle, and the electric fields are equal by solving the constraints. Now we cut the circle into two regions. Then electric fields become the non-trivial centers when we remove the spatial Wilson line. Because the entangling surface in the two-dimensional Abelian gauge theory is a point, we cannot define the non-local Wilson line on the entangling surface. Therefore, the centers live in the bulk region. In this case, we cannot just remove operators from the entangling surface. Therefore, the mutual information possibly might be different from different observation methods. We argue that if the mutual information is independent of the observation methods, we cannot find the non-trivial centers from removing operators in the two-dimensional Abelian gauge theory.

\subsection{First-Order Formulation}
The quantization algebra and constraint in the three dimensional Chern-Simons gauge theory are given by the following:
\bea
&&\lbrack A_1^a(x), A_2^b (y)\rbrack=\delta^{ab}\delta(x-y), \qquad \lbrack A_1^a(x), A_1^b(y)\rbrack=0, 
\nn\\
&& \lbrack A_2^a(x), A^b_2(y)\rbrack=0 , \qquad F_{12}^a=0.
\eea
We again choose the gauge fixing condition $A_0^a=0$ and $\partial_1A^{1, a}+\partial_2A^{2, a}=0$. Combining with the constraints, and then one obtains that
the gauge field on an entangling surface only depends on the time after one of its component is removed. This implies that one cannot remove operators to obtain non-trivial centers in three dimensional Chern-Simons gauge theory with a trivial topology.
\\

We reach the same conclusion through a different route. The only physical operators in the three-dimensional Chern-Simons gauge theory is the Wilson loop. For the Wilson loops of contractible cycles, we obtain the vanishing field strength $F_{\mu\nu}=0$. In other words, there are no local operators in the three-dimensional Chern-Simons theory on space with a trivial topology. 
\\

In the case of the non-trivial topology, we consider a spatial two torus in the three-dimensional Chern-Simons gauge theory. We have two different spatial Wilson lines wrapping the $A$-cycle and $B$-cycle. The $A$-cycle is like an electric field and $B$-cycle is like a spatial Wilson line. Then the Wilson line wrapping the $A$-cycle can be a non-trivial center when we remove the Wilson line wrapping the $B$-cycle. In this case, the centers also live in the bulk region. Hence we also find that if the mutual information is independent of the observational methods, we do not find the non-trivial centers in the three-dimensional Chern-Simons gauge theory.
\\

We also find that the situation of two-dimensional Yang-Mills gauge theory is the same as in the three-dimensional Chern-Simons gauge theory. Hence the reason should be that removing non-dynamical operators do not have any effect on the entanglement on the trivial topology, and it may be hard to create non-trivial centers, which only lives on the entangling surface.
\\

Finally, we discuss the fermion theory. The quantization algebra is given by:
\bea
&& \{\psi_i^{\dagger}(x), \psi_j(y)\}=-i\delta_{ij}\delta(x-y), \qquad   \{\psi_i(x), \psi_j(y)\}=0, 
\nn\\ 
&& \{\psi_i^{\dagger}(x), \psi_j^{\dagger}(y)\}=0.
\eea
If we remove the fermionic field $\psi_i$ on an entangling surface, and then it appears that $\psi_i^{\dagger}$ is also removed. Therefore, we do not seem to have any non-trivial center in the fermion theory either. This result is mainly due to the particular form of quantization algebra in the first-order formulation. Hence we argue that any system with the first-order formulation does not have non-trivial centers from removing operators of an entangling surface. We also suspect that topological theory does not have non-trivial centers on an entangling surface.

\section{The Entanglement with Centers in the $p$-Form Non-Interacting Theory}
\label{Sec:3}
We move on to the entanglement with centers in the $p$-form non-interacting quantum field theory. From our computation, we can gain some insights into two problems. The first problem is whether one obtains contributions to a universal term of the entanglement entropy from the boundary term of the on-shell action in non-gauge theory. The second problem is whether introducing a boundary term is equivalent to doing a regularization for the zero-modes of the cone directions. Because only the non-interacting quantum field theory is considered, we separately discuss the bulk entanglement and boundary entanglement. In the non-interacting quantum field theory, we find a suitable form of strong subadditivity with centers. We first review the results of boundary entanglement entropy in the Abelian one-form gauge theory \cite{Donnelly:2015hxa, Donnelly:2014fua}, and then we extend the study to the massive non-interacting scalar field theory and the $p$-form Abelian gauge theory. Finally, we discuss the strong subadditivity in the non-interacting theory. The simple review of Lagrangian formulation for the entanglement with centers in Appendix \ref{app2} and the simple example for the one-form gauge theory in Appendix \ref{app3}. We provide the details of the bulk entanglement entropy in the Abelian $p$-form non-interacting theory in Appendix \ref{app4} and show the relation between the sphere manifold and two-dimensional cone in Appendix \ref{app5}.

\subsection{Boundary Entanglement Entropy in\\ the Massive Non-Interacting Scalar Field Theory}
We use a similar method to analyze an on-shell boundary action in the massive non-interacting scalar field theory. The equation of motion of the massive non-interacting scalar field theory  is
\bea
\bigg(\nabla_{\mu}\nabla^{\mu}+m^2\bigg)\phi=0,
\eea
and we solve the equations of motion from the spacetime interval
\bea
ds^2=dr^2+r^2d\theta^2+dx_{\bot}^2.
\eea
The solution of the scalar field is the following
\bea
\phi=\sum_{n^{\prime}}\phi_{n^{\prime}}(r)\psi_{n^{\prime}}(x_{\bot}),
\eea
where
\bea
\nabla^2\psi_{n^{\prime}}(x_{\bot})=-\lambda_{n^{\prime}}\psi_{n^{\prime}}(x_{\bot}).
\eea
Plugging the solution of the scalar field into the equation of motion, we obtain the following:
\bea
&&\sum_{n^{\prime}}\bigg\lbrack\bigg(\partial_r\partial^r\phi_{n^{\prime}}(r)\bigg)\psi_{n^{\prime}}+\phi_{n^{\prime}}(r)\partial_{x_{\bot}}^2\psi_{n^{\prime}}
\nn\\
&&+\Gamma^{\theta}{}_{\theta r}\partial^r\phi_{n^{\prime}}\psi_{n^{\prime}}(x_{\bot})\bigg\rbrack+m^2\phi
\nn\\
&=&\sum_{n^{\prime}}\bigg(\partial_r^2\phi_{n^{\prime}}-\lambda_{n^{\prime}}\phi_{n^{\prime}}+\frac{1}{r}\partial_r\phi_{n^{\prime}}\bigg)\psi_{n^{\prime}}+m^2\phi
\nn\\
&=&\sum_{n^{\prime}}\bigg(\partial_r^2\phi_{n^{\prime}}+\frac{1}{r}\partial_r\phi_{n^{\prime}}-\lambda_{n^{\prime}}\phi_{n^{\prime}}\bigg)\psi_{n^{\prime}}
+m^2\phi
\nn\\
&=&0.
\eea
Therefore, the equation for $\phi_{n'}$ follows as that
\bea
\partial_r^2\phi_{n^{\prime}}+\frac{1}{r}\partial_r\phi_{n^{\prime}}+(m^2-\lambda_{n^{\prime}})\phi_{n^{\prime}}=0.
\eea
Because the entangling surface is set at $r=0$, and $\phi$ should not be singular near the entangling surface, the asymptotic behavior of the scalar field follows from that
\bea
2\partial_r^2\phi_{n^{\prime}}+(m^2-\lambda_{n^{\prime}})\phi_{n^{\prime}}=0,
\eea
in which we assumed that $\partial_r\phi_n\sim 0$ near the entangling surface to avoid the singularity.
Therefore, the solution near the entangling surface is given by
\bea
\phi_n\sim a_n\cos\bigg(\sqrt{\frac{1}{2}(m^2-\lambda_n)}r\bigg),
\eea
where $a_n$ are arbitrary constants. 
\\

This result is interesting because the result implies that a universal term of the entanglement entropy does not receive contributions from the on-shell boundary term in the case of the massive non-interacting scalar field theory with a planar entangling surface because of $\partial_r\phi_n\sim 0$ near the boundary. The ambiguity persists in this case although we do not find the universal contributions of entanglement entropy from the boundary on-shell action. In this case, all values of centers have the same weight. Therefore, the choice of non-trivial centers in the massive non-interacting scalar field theory does not lead to any universal contribution of entanglement entropy. We also give a quick comment to the interacting scalar field theories (only local interacting terms). If we use the same method to analyze local interacting scalar field theory, we still have $\partial_r\phi_n\sim 0$ near the entangling surface. Therefore, the result is the same as in the massive non-interacting scalar field theory. In other words, a direct sum decomposition of Hilbert space in the scalar field theory possibly does not give any interesting results. We already showed that the fermion theory does not have the choice of non-trivial centers by removing operators. Hence gauge symmetry may connect bulk and boundary sides to allow a universal contribution to the entanglement entropy from the on-shell boundary action in gauge theory.

\subsection{Boundary Entanglement Entropy in the Abelian $p$-Form Gauge Theory}
We generalize the computation of the Abelian one-form gauge theory to the case of Abelian $p$-form gauge theory. The equation of motion in the Abelian $p$-form gauge theory is
\bea
\nabla_{\mu_1}F^{\mu_1\mu_2\cdots\mu_{p+1}}=0.
\eea
As before, we also look for the asymptotic solution of equation of motion. To analyze the asymptotic behavior of equation of motion easily, we choose the solution
\bea
A_{\theta x_{\bot}}=\sum_{n^{\prime}}\phi_{n^{\prime}}(r)\psi_{n^{\prime}}(x_{\bot}),
\eea
where $\nabla_{x_{\bot}}^2\psi_{n^{\prime}}=-\lambda_{n^{\prime}}\psi_{n^{\prime}}$, and other components are zero, and $A_{\theta x_{\bot}}$ is a $p$-form field. The transverse space $x_{\bot}$ are labeled by the multiple indices. The equation of motion reads:
\bea
&&\nabla_{\mu_1}F^{\mu_1\mu_2\cdots\mu_{p+1}}
\nn\\
&=&
\partial_{\mu_1}F^{\mu_1\mu_2\cdots\mu_{p+1}}+\Gamma^{\mu_1}{}_{\nu\mu_1}
F^{\nu\mu_2\mu_3\cdots\mu_{p+1}}=0,
\eea
when $p> 0$. The non-trivial part of the equation of motion gives
\bea
\partial_rF^{r\theta x_{\bot}}+\partial_{x_{\bot}^{\prime}}F^{x_{\bot}^{\prime}\theta x_{\bot}}+\Gamma^{\theta}{}_{r\theta}F^{r\theta x_{\bot}}=0.
\eea
Therefore, we get
\bea
\sum_{n^{\prime}}\bigg\lbrack\partial_r\bigg(\frac{1}{r^2}\partial_r\phi_{n^{\prime}}\bigg)+\frac{1}{r^3}\partial_r\phi_{n^{\prime}}-\frac{1}{r^2}\lambda_{n^{\prime}}\phi_{n^{\prime}}\bigg\rbrack\psi_{n^{\prime}}=0
\nn\\
\eea
and obtain
\bea
\partial_r^2\phi_{n^{\prime}}-\frac{1}{r}\partial_r\phi_{n^{\prime}}-\lambda_{n^{\prime}}\phi_{n^{\prime}}=0.
\eea
Hence the boundary partition function of $p$-form non-interacting theory is:
\bea
Z^\zt{BOPAG}&\sim&\prod_{n^{\prime\prime}}\bigg(-\frac{\ln(\epsilon^{-1})\lambda_{n^{\prime\prime}}}{\beta}\bigg)^{\frac{1}{2}}
\nn\\
&=&\det{}^{\prime} \bigg(\frac{\ln(\epsilon^{-1})\nabla_{x_{\bot}}^2}{\beta}\bigg)^{\frac{1}{2}},
\eea
where as before we exclude $\lambda_{n^{\prime \prime}}=0$ in the product. The partition function gives the boundary entanglement entropy through the conical method.

\subsection{Bulk Entanglement Entropy in the Massive Non-Interacting Scalar Field Theory}
We use the conical method to compute the entanglement entropy as
\bea
S_{EE}=\bigg(1-\beta\frac{\partial}{\partial\beta}\bigg)\ln Z(\beta)\bigg|_{\beta=2\pi}.
\eea
Therefore, the entanglement entropy is obtained from the partition function. The Lagrangian of non-interacting scalar field theory is
\bea
S_{SF}=\int d^Dx\ \sqrt{\det g_{\nu\rho}}\ \bigg( \frac{1}{2}\nabla_{\mu}\phi\nabla^{\mu}\phi+\frac{m^2}{2}\phi^2\bigg).
\eea
The bulk term of action after dropping all the boundary terms is
\bea
\int d^Dx\ \sqrt{\det g_{\nu\rho}}\ \bigg\lbrack \frac{1}{2}\phi\bigg(-\nabla_{\mu}\nabla^{\mu}+m^2\bigg)\phi\bigg\rbrack.
\eea
Therefore, the partition function in bulk is determined by
\bea
\bigg(\det\big(-\Box+m^2\big)\bigg)^{-\frac{1}{2}},
\eea
up to a normalization constant. Hence the free energy of the non-interacting scalar field theory is
\bea
&&\frac{1}{2}\ln\det\big(-\Box+m^2\big)
\nn\\
&\sim&-\frac{1}{2}\int d^Dx\sqrt{\det g_{\mu\nu}}\int_{\epsilon^2}^{\infty}\frac{ds}{s}\ e^{-sm^2}K(s, x, x),
\eea
where
\bea
K(s, x, x^{\prime})=\langle x\mid e^{-s(-\Box)}\mid x^{\prime}\rangle.
\eea
\\

After we compute the bulk free energy, we could obtain the bulk entanglement entropy by the conical method. For bulk entanglement entropy, please see Appendix \ref{app4}.

\subsection{Universal Term of Entanglement Entropy}
We discuss the universal term of entanglement entropy in the $p$-form non-interacting theory. We compute the entanglement entropy in the Euclidean flat background with an $S^{2p}$ entangling surface. In this case, the universal term of entanglement entropy in the $p$-form non-interacting theory can be expressed in terms of the universal terms of massless non-interacting scalar field theory in various even dimensions. We also give the results for the universal terms of entanglement entropy in the case of the $p$-form non-interacting theory in $p+1$ and $p+2$ dimensions.
\\

In Appendix \ref{app5}, we review why the computation of the entanglement entropy in the Euclidean polar coordinate with the entangling surface $S^{D-2}$ is equivalent to the case of $S^D$ with a unit radius in conformal field theory \cite{Casini:2011kv}. In Appendix \ref{app5}, we also interpret why introducing boundary conditions to a two-dimensional cone is equivalent to computing the boundary entanglement entropy on a sphere manifold.
\\

To obtain the universal term of entanglement entropy in the $p$-form non-interacting theory in $2p+2$ dimensions, we need to rewrite the heat kernel of $p$-form non-interacting theory in terms of those theories of the lower forms. Because the $p$-form field has a ($p-1$)-form gauge parameter, we need to cancel the ($p-1$)-form degrees of freedom. Since the ($p-1$)-form field also has its own gauge symmetry, we also need to cancel ($p-2$)-form degrees of freedom. The procedure continues until we meet scalar fields. Therefore, the heat kernel of $p$-form becomes the following term
\bea
&&2K^{2p+2}_{p-1}-3K^{2p+2}_{p-2}+\cdots
\nn\\
&&+(-1)^{p}(p+1)K^{2p+2}_0+C^{2p}_{p}K_0^{2p+2},
\eea
which comes from $K_p^{2p+2}$ on the bulk up to zero modes, where $K_p^q$ is the heat kernel of the $p$-form on $S^q$. The contribution of the ghost fields in the
bulk is given by
\bea
&&-2K^{2p+2}_{p-1}+3K^{2p+2}_{p-2}+\cdots
\nn\\
&&+(-1)^{p+1}(p+1)K^{2p+2}_0.
\eea
We need to add the zero modes of two-dimensional cone directions, which come from the boundary terms, and subtract the zero modes of transverse directions. The contribution of zero modes of the two-dimensional cone directions to the heat kernel of $p$-form non-interacting theory is
\bea
&&-2C^{2p-2}_{p-1}K_0^{2p}+3C^{2p-4}_{p-2}K_0^{2p-2}+\cdots 
\nn\\
&&+(-1)^p(p+1)K_0^2.
\eea
Hence the total heat kernel of $p$-form non-interacting theory is 
\bea
\label{uni_total}
&&C^{2p}_pK_0^{2p+2}-2C^{2p-2}_{p-1}K_0^{2p}+\cdots 
\nn\\
&&+(-1)^p(p+1)K_0^2.
\eea 
We use the universal term of entanglement entropy in the massless scalar field theory in various even dimensions to determine the universal term of entanglement entropy in $p$-from non-interacting theory in $2p+2$ dimensions. The universal term of entanglement entropy in the $p$-form non-interacting theory in $2p+2$ dimensions as the following
\bea
&&C_p^{2p}
\nn\\
&&\times\bigg(\mbox{universal terms of the 0-form non-interacting theory}
\nn\\
&&\mbox{ on $S^{2p}$}\bigg)
\nn\\
&&-2\bigg(\mbox{universal terms of} 
\nn\\
&&\mbox{the $(p-1)$-form non-interacting theory on $S^{2p}$}\bigg)
\nn\\
&&-\bigg(\mbox{universal terms of}
\nn\\
&&\mbox{ the $(p-2)$-form non-interacting theory on $S^{2p-2}$}\bigg).
\nn\\
\eea
The universal terms of entanglement entropy in the massless scalar field theory on an even-dimensional sphere was already computed \cite{Casini:2010kt}. The boundary contribution to the universal term of $p$-form non-interacting theory (on $S^{2p+2}$) is of opposite sign to that of the ($p-1$)-form non-interacting theory (on $S^{2p}$). Therefore, we determine the universal term of bulk entanglement entropy in the $p$-form non-interacting theory in $2p+2$ dimensions as
\bea
&&C_p^{2p}
\nn\\
&&\times\bigg(\mbox{universal terms of}
\nn\\
&&\mbox{ the 0-form non-interacting theory on $S^{2p}$}\bigg)
\nn\\
&&-\bigg(\mbox{universal terms of}
\nn\\
&&\mbox{ the $(p-1)$-form non-interacting theory on $S^{2p}$}\bigg)
\nn\\
&&-\bigg(\mbox{universal terms of the $(p-2)$-form free theory}
\nn\\
&&\mbox{ on $S^{2p-2}$}\bigg).
\eea
The universal term of entanglement entropy \er{uni_total} is also consistent with the known anomaly coefficients \cite{Cappelli:2000fe} (see also \eg Ref. \cite{CamporesiHiguchi} for the computation of bulk part).
\\

From our computation methods, it is very easy to determine the universal term of entanglement entropy in the $p$-form non-interacting theory in $p+1$ and $p+2$ dimensions. In the case of $p+1$ dimensions, we do not have the dynamical degrees of freedom. Therefore, the universal term of entanglement entropy vanishes. In the case of the $p$-form non-interacting theory in $p+2$ dimensions, the universal term of boundary entanglement entropy is determined by the ($p-1$)-form non-interacting theory in $p$ dimensions. Therefore, the universal term of boundary entanglement entropy vanishes in the $p$-form non-interacting theory in $p+2$ dimensions. Then the $p$-form non-interacting theory in $p+2$ dimensions is dual to the 0-form non-interacting theory in $p+2$ dimensions. Therefore, we know that the universal term of entanglement entropy in the $p+2$ dimensional $p$-form non-interacting theory is the same as that of the 0-form non-interacting theory in $p+2$ dimensions.

\subsection{Strong Subadditivity in the Non-Interacting Theory}
The strong subadditivity \cite{Araki:1970ba, Lieb:1973cp} is satisfied generically if three algebras are mutually commuting. In the case of non-trivial centers, the strong subadditivity is not satisfied generically \cite{Ma:2015xes, Casini:2014aia, VanAcoleyen:2015ccp} because of the losing mutually commuting three algebras. In the case of the non-interacting scalar field theory, the centers have an equal probability distribution. Then it is easy to show that the strong subadditivity holds in the non-interacting scalar field theory. Even if we consider the equal probability distribution of centers in the interacting scalar field theory, the strong subadditivity may not hold. 
\\

We first use the inequality
\bea
\mathrm{Tr}\bigg(A\ln A-A\ln B-A+B\bigg)\ge 0
\eea
 with $A=\rho_{123}$ and $B=\exp(-\ln\rho_2+\ln\rho_{12}+\ln\rho_{23})$ to find the inequality:
\bea
&&F(\rho_{123})
\nn\\
&=&S_{123}+S_2-S_{12}-S_{23}
\nn\\
&\le&\mbox{Tr}\bigg(\exp(\ln\rho_{12}-\ln\rho_2+\ln\rho_{23})-\rho_{123}\bigg),
\eea
and apply the inequality
\bea
\mathrm{Tr}\bigg(e^CT_{\exp(-A)}\big(e^B\big)\bigg)\ge\mathrm{Tr}\bigg(e^{A+B+C}\bigg),
\eea
where
\bea
\frac{d}{dx}\ln(\alpha+x\beta)|_{x=0}&\equiv& T_{\alpha}(\beta)
\nn\\
&\equiv& \int_0^{\infty} dy\ (\alpha+y1)^{-1}\beta(\alpha+y1)^{-1},
\nn\\
\eea
  to obtain the inequality
\bea
&&\mbox{Tr}\bigg(\exp(\ln C-\ln D+\ln E)\bigg)
\nn\\
&\le&\mbox{Tr}\bigg(\int_0^{\infty}dx\ C(D+x1)^{-1}E(D+x1)^{-1}\bigg).
\eea
Hence we get the inequality:
\bea
&&F(\rho_{123})
\nn\\
&\le&\mbox{Tr}\bigg(-\rho_{123}+\int_0^{\infty}dx\ \rho_{12}(\rho_2+x1)^{-1}\rho_{23}(\rho_2+x1)^{-1}\bigg)
\nn\\
&=&-\mbox{Tr}\ \rho_{123}+\mbox{Tr} \bigg(\int_0^{\infty}dx\ \rho_2(\rho_2+x1)^{-1}\rho_2(\rho_2+x1)^{-1}\bigg)
\nn\\
&=&\mbox{Tr}\ \rho_2-\mbox{Tr}\ \rho_{123}
\nn\\
&=&0,
\eea
in which we use $C=\rho_{12}$, $D=\rho_2$, and $E=\rho_{23}$. The reduced density matrix is $\rho=\oplus_j p_j\rho_j$, where $p_j$ is the probability distribution of centers, and $\rho_j$ depends on the choice of centers. Notice that we implicitly assume that $\rho_1, \rho_2, \rho_{23}, \cdots$ can be obtained by a partial trace operation from the density matrix $\rho_{123}$. The sufficient condition is that we have three sets of centers (associated with each region) that commute with each other \footnote{In principle, one can choose different centers in the reduced density matrix $\rho_1$ (rather than the one inherited from the density matrix $\rho_{123}$), and the strong subadditivity is likely to be violated, but we deem this kind of definition to be physically uninteresting.}. 
\\

To see how this scenario works explicitly, we turn to the non-interacting theory, in which each reduced density matrix $\rho_j$ gives the same entanglement entropy. In the non-interacting theory, we can use a reduced density matrix with the probability distribution of centers to describe the entanglement in different regions. Therefore, we show that the strong subadditivity remains valid in the non-interacting theory. It is not clear to us whether the proof can be extended to the generic interacting theory due to the lack of proof about the existence of mutually commuting centers in different regions. 
\\

The above proof shows that we also have the strong subadditivity in the Abelian gauge theory. If we consider the Abelian gauge theory on a lattice, the entanglement in this theory is hard to be described by a reduced density matrix with a probability distribution. Therefore, the strong subadditivity is not satisfied generically \cite{Casini:2014aia, VanAcoleyen:2015ccp} in the case of the finite lattice spacing. The violation of the strong subadditivity in the lattice Abelian gauge theory is not in contradiction with the result at the continuum limit above. 
\\

Finally, we rewrite the strong subadditivity in a different form. From $S_{123}=S_4$ and $S_{12}=S_{34}$, the strong subadditivity is
\bea
S_4+S_2\le S_{34}+S_{23}.
\eea
Therefore, we can rewrite the strong subadditivity as the following
\bea
S_1+S_2\le S_{13}+S_{23}.
\eea
We remind the reader that the strong subadditivity is satisfied due to the fact that the entanglement entropy is the sum of bulk entanglement entropy and classical Shannon entropy. Therefore, this proof cannot be extended to interacting theories in general. 

\section{Mutual Information with Centers in CFT$_2$}
\label{Sec:4}
To analyze the
effects of the choice of centers in mutual information, we consider disjoint regions
and the tensor product decomposition of Hilbert space, but we generate the non-trivial centers by removing some operators in each region. We consider a planar case in the massive non-interacting scalar field theory to show that mutual information does not depend on the choice of centers. In the case of CFT$_2$, we calculate the mutual information of multiple intervals. The result also provides supporting evidence that the mutual information does not depend on the choice of centers.
\\

In CFT$_2$, the entanglement entropy can be computed by the replica trick. One way is to take the field on different sheets as different fields (\ie working in CFT$^n$/$Z_n$) and introduce twist operators. We then do the computation (of $n$-point functions of twist operators)
on a sphere. Here we are more interested in the other approach, in which
one performs the path integral in a covering space. The conical singularity is usually taken care of by cutting off the tip of the two-dimensional cone and gluing back a disk \eg Ref. \cite{Faulkner:2013yia}. This is essentially the smooth cone regularization (see \eg Ref. \cite{Lewkowycz:2013laa}). Alternately,
one can impose a boundary condition on the little circle of radius $\xe$ around the tip of the two-dimensional cone, which creates the boundary state. The boundary state for the smooth cone prescription follows from the insertion of the identity operator. In
principle, we can choose other boundary states, and hence the ambiguity of entanglement entropy arises. 
\begin{figure}
\includegraphics[width=1.\textwidth]{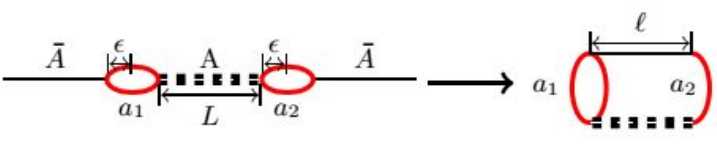}
\caption{The path integral representation of the reduced density matrix in the region $A$, $\rho_A$, under conformal transformations.   
                In two dimensions, the entangling surface consists of two points, which become two tiny circles after imposing the cutoff. We can specify the different boundary conditions, $a_1, a_2$, on the boundary circles.
}
\label{fig:2cut}
\end{figure}
\\
     
To see how the boundary conditions affect the entanglement entropy, let us consider a single interval with the length $L$ \cite{Ohmori:2014eia}. For computing the R\'enyi entropy $S_n$
using the replica trick, the two conical singularities (endpoints of the interval at the points $z_1$ and $z_2$) are removed and some boundary conditions $a_{1,2}^{(n)}$ are imposed on the little circle of the radius $\xe$. The conformal mapping of the form of 
\be
w = \ln \frac {z - z_1}{z-z_2}\,
\ee
gives a cylinder of the circumference $2\pi$ and length $\ell =  \ln \lb 
L /\xe\rb^{2}$. The partition function then reads
\bea
        &&Z_n
        \nn\\
       &=&\bigg\langle a_1^{(n)}\bigg| \exp\bigg\lbrack\frac{\ell}{n} \bigg(\frac{c+\tilde{c}}{24}- L_0-\tilde{L}_0\bigg)\bigg\rbrack \bigg|a_2^{(n)}\bigg\rangle \label{eq:Zn}\,,
\eea
where $|a_{1,2}^{(n)}\rangle$ are the boundary states from the cutoff circle. We insert a complete set of states as intermediate states,
\bea
\el{cylinderamp}
&&Z_n 
\nn\\
&=& 
\langle a_1^{(n)}\st{0} \exp\left(\frac{\ell}{n}\frac{c}{12}\right) \langle 0\st{a_2^{(n)}}  
\nn\\
&&+ \langle a_1^{(n)}\st{\mathcal{O}}\exp\bigg\lbrack\frac{\ell}{n}\bigg(\frac{c}{12}-\Delta_\CO\bigg)\bigg\rbrack \langle\CO \st{a_2^{(n)}} +\cdots.
\nn\\
\eea
The R\'enyi entropy is computed using the below way:
\be
\el{Renyi}
S_n = \frac {\ln \Tr \rho^n} {1-n} =\frac {\ln Z_n - n \ln Z_1} {1-n}\,,
\ee
has the following the expansion in terms of $L/\xe$ as that
\bea
        &&S_n 
        \nn\\
        &= &\bigg(1+\frac1n\bigg)\frac{c}{6}\ln\frac{L}{\epsilon} 
        \nn\\
        &&+ \frac{1}{1-n}\bigg(s(a_1^{(n)})-ns(a_1^{(1)}) + s^*(a_2^{(n)})-ns^*(a_2^{(1)})\bigg) 
        \nn\\
        && +  \frac{1}{1-n}
        \frac{\langle a_1^{(n)}\st{\mathcal{O}}\langle \mathcal{O}\st{a_2^{(n)}}}{\langle a_1^{(n)}\st{0}\langle 0\st{a_2^{(n)}}}
        \left(
        \frac{L}{\epsilon}
        \right)^{-2\Delta_{\mathcal{O}}/n}
        +\cdots\,,        \label{eq:subleadingRenyi}
\eea
where $s(a_1^{(n)}) = \ln \langle {a_1^{(n)}}\st 0$ is the boundary entropy. Because the boundary entropy may not be a real number, the R\'enyi entropy has the complex conjugate of boundary entropy. It is easy to see the leading term $\big((1+1/n) c/6\big) \ln(L/\epsilon)$ of R\'enyi entropy agrees with our familiar result. The rest depends on the choice of the boundary condition $a_{1,2}^{(n)}$, and hence is ambiguous to the entanglement entropy. 
\\

An explicit form the entanglement entropy can be obtained from the R\'enyi entropy by taking the limit $n\rightarrow 1$. The finite and real entanglement entropy depends on the choice of quantum states. There must be some constraints in what
states can be inserted as the boundary states $\st{a_{1,2}^{(1)}}$. 
\\

We obtain some features of the boundary states $\st{a_{1,2}^{(n)}}$
by comparing a cylinder amplitude with that on a sphere. According to a
state-operator mapping, a vertex operator inserted in the past or future infinity, that is $z=0$ or $z=\infty$ can be considered as the initial or final states. The inner product is just the path integral with two insertions at the two positions, $0, \infty$ as below
\bea
\label{2Pf}
\langle \langle i|j\rangle= \langle {\mathscr{A'}}_i (\infty,\infty) {\mathscr{A}}_j (0,0) \rangle_{S_2}.
\eea
The prime operator is defined in the $u$-frame at the other pole ($u=1/z$)
and hence is related to the unprimed by the following
\bea
\mathscr{A'}{}_i (z,\bar z) = z^{-2h_i} {\bar z}^{-2{\tilde h}_i}\mathscr{A}{}_i (z,\bar z)\,.
\eea
The dual state $\langle \langle i|$ is not the conjugate of the state $\st{i}$, and they are
different from a finite normalization factor:
\bea
\label{normfactor}
\langle \langle i| = K \langle i|,\qquad K = i C_{S_2}\,,
\eea
where $C_{S_2}$ is the vacuum partition function on the sphere. Now a conformal
mapping to a cylinder gives a partition function that can be interpreted as the following
\begin{equation}
        Z= \bigg\langle \tilde i| \exp\lsb\ell \lb \frac{c+\tilde{c}}{24}- L_0-\tilde{L}_0\rb\rsb \bigg|\tilde j\bigg\rangle \,.
\end{equation}
The factor $\exp\bigg( (c+\tilde{c})\ell/24\bigg)$ can be understood as
rising from the conformal anomaly. To make the rest agrees with the ground state of sphere
partition function, we need to take the quantum state
\bea
\st{\tilde j} = \exp \bigg(\frac \ell 2 L_0+\frac \ell 2 \tilde L_0\bigg)\st{j}\,,
\eea
which is essentially the evolution of a state from $\tau=0$ ($|z| = 1$) to
$\tau = -l/2$. This is consistent with the time evolution
in the radial quantization because the time revolution is generated by $\exp\big(-(L_0+\tilde
L_0)(\tau_2 - \tau_1)\big)$. Notice that the exponential factor blows up by taking the limit $\ell
\to \infty$. If we want to obtain finite and real entanglement entropy, it is necessary to have a constraint in the boundary state.
\\

In fact, the above case implies that the boundary condition is conformal invariant
(at least in the limit $\xe \to 0$). To see what precisely this means,
one can consider the inner product between the state determined by some boundary conditions (specified by some field configurations $\phi_\xe$) and the reference state $\langle\langle\CO_i|$. This can be expressed as the partition function $Z_\xe$ on a sphere with the operator $\CO_i$ at the position $u=0$ and a boundary condition at the position $|z|=\xe$. Now we move to a new boundary at the position $\xe' =
\xl \xe$. To compare with the previous case, one can perform the conformal transformation $z' = \iv \xl z$ and
$u' = \xl u$ so that the transition function remains invariant as the relation $z' u' = 1$.
The insertion of the primary operator $\CO_i$ becomes $\xl^{-h_i} \bar \xl^{-\tilde h_i} \CO_i'$ in the new coordinate. Imposing the same boundary condition $\phi_\xe$ and the total partition function $Z_{\xl \xe}$, which is invariant under the conformal transformation as the below
\bea\el{diffbnd} Z_{\xl \xe} = \xl^{-h_i} \bar \xl^{-\tilde h_i} Z_{\xe}
\eea
up to the conformal anomaly.
Switching back to the cylinder, each partition function can be expressed as the following terms:
\bea
Z_{\xl \xe} &=& \exp\lsb(\ell+\ln \xl)\lb\frac{c+\tilde{c}}{24}- h_i-\tilde h_i\rb\rsb \langle\langle i \st{\phi_{\xl \xe}},
\nn\\
 Z_{\xe} &=& \exp\lsb\ell \lb \frac{c+\tilde{c}}{24}- h_i-\tilde h_i\rb\rsb \langle\langle i \st{\phi_{\xe}}\,.
\eea
This computation is done without doing the conformal transformation $z' = \iv \xl z$ and the extra factor $\ln \xl$ in $Z_{\xl \xe}$ is due to the difference on the boundary (propagation from the point $\ln( \xl \xe)$ to infinity $+\infty$ instead of from the position $\ln \xe$).
This extra factor is precisely the same as the difference in eq. \er{diffbnd} (there is another $c$-dependent factor, which follows from the conformal anomaly). As a result, we have 
\bea
\langle\langle i\st{\phi_{\xl \xe}} = \langle\langle i \st{\phi_{\xe}}
\eea 
if the boundary configuration $\phi_{\xl \xe}$ (at $|z| = \xl \xe$) that defines that the state $\st{\phi_{\xl \xe}}$ is obtained from the conformal transformation of the configuration $\phi_{\xe}$ (at the position $|z| = \xe$). In other words, the boundary condition $\phi_{\xe \to 0}$ follows from the dilation of some arbitrary boundary conditions at the place $|z| = 1$, which seems to be a quite natural way for imposing a boundary condition. Nevertheless, the massless non-interacting scalar field theory is an example to know the existence of finite and real boundary entropy.
\\

Ignoring all the issues and simply playing with their formula, we obtain
some interesting results. For example, we compute the mutual information
$I(A^{\prime}:B^{\prime}) = S(A^{\prime}) + S(B^{\prime})- S(A^{\prime} \cup B^{\prime})$ with $A^{\prime}: z< z_1,\, B^{\prime}: z> z_2$. Computation of the boundary entropy in the entanglement entropy in the region $A^{\prime}\cup B^{\prime}$, $S(A^{\prime} \cup B^{\prime})$, is the same as computing the boundary entropy in an interval between the position $z_2$ and the position $z_1$ or can be obtained by computing the entanglement entropy on two intervals, $z<z_1$ and $z>z_2$, from the similar way of the single interval. Then the entanglement entropy in the region $A$, $S(A^{\prime})$, and the entanglement entropy in the region $B^{\prime}$, $S(B^{\prime})$, can be obtained
by sending the limit $L/\xe \to \infty$. In the limit $\xe \to 0$, the only ambiguous
contribution follows from the boundary entropy, which is local. Moreover, the entanglement entropy in the region $A^{\prime}$,
$S(A^{\prime})$, and the entanglement entropy in the region $B^{\prime}$, $S(B^{\prime})$, vanishes if the vacuum at infinity (there is
no boundary essentially). Hence we obtain that the ambiguous terms are canceled in the mutual information, and the mutual information is unambiguous as expected.

\subsubsection{Mutual Information of Multiple Intervals}
The mutual R\'enyi information in the region $A$ for $N$ intervals 
\be
A = \bigcup_{i=1}^N A_i = [z_1, z_{2}] \cup  [z_3, z_4 ] \ldots \cup [z_{2N-1},z_{2N}]\,
\ee
can be computed from a partition function of a higher genus surface $\CM$ \cite{Faulkner:2013yia, Headrick:2012fk}. 
\\

Different boundary conditions are realized by replacing the identity operator (in the case of a smooth cone) with other states. The partition function of a higher
genus surface is then computed with an appropriate boundary condition. We put the boundary at the place $|z|=\xe$ of some local coordinate systems, then consider the cutting over the region $|z|=1$, and also insert a complete set of the operators (states) $\CO_i$, which turns the partition function on the manifold $\CM$ to the below form
\bea
\el{SCCFT:1} 
&&\int_\CM (\dots_1)e^{-S[\phi]}[d\phi]{\Psi(\phi_\xe)}
\nn\\
& =& \sum_{ij} \langle \dots_1 \CO_i\ke_{\CM} G^{ij} \int_{S^2}\CO_j e^{-S[\phi]}[d\phi]{\Psi(\phi_\xe)}\,,
\eea
where $\Psi$ is a wave function determined by the boundary condition at the region $|z|=\xe$. We further perform the mapping $w = \ln z$ to take the sphere into a long cylinder with the length $-\ln \xe^2$. Therefore, we obtain the conclusion for that only the contribution from the operator $\CO_j = 1$, which survives in the limit $\xe \to 0$, if we only consider a unique vacuum state, and hence the difference is due that replacing the boundary condition by the identity operator is just an extra term that corresponds to the boundary entropy. This surgery procedure can be performed locally for all the end points $z_i$ of the intervals. As a result, the R\'enyi entropy $S_n[A;\{a_i^{(n)}\}]$ in the region $A$ that consists of multiple intervals gives the below results
\bea
\label{eq:multiRenyi} 
&&S_n[A;\{a_i^{(n)}\}] 
\nn\\
&=& S_n[A;\{1\}] 
\nn\\
&&+ \sum_{i=1}^{N} \frac{1}{1-n}
\nn\\
&&\times\bigg(s(a_{2i-1}^{(n)})-ns(a_{2i-1}^{(1)})+s^*(a_{2i}^{(n)})-ns^*(a_{2i}^{(1)})\bigg) 
\nn\\
&&+ \CO(\xe),
\eea
which is the generalization of eq. \er{eq:subleadingRenyi} to multi-interval. The notation $\{a_i^{(n)}\}$ denotes the boundary conditions at various points $z_i$, and $\{1\}$ means inserting identity everywhere. The extra contribution due to boundary condition is given by the following
\bea
s(a_i^{(n)}) = \ln \langle {a_i^{(n)}}\st 0.
\eea
Now we replace each term in the R\'enyi mutual information
\bea
I_n(A^{\prime}:B^{\prime}) = S_n (A^{\prime}) + S_n(B^{\prime}) - S_n(A^{\prime}\cup B^{\prime})
\eea
by eq. \er{eq:multiRenyi} and get the following result
\bea 
&&I_n[A^{\prime}:B^{\prime};\{a_i^{(n)}\}] 
\nn\\
&=& S_n[A^{\prime};\{1\}] + S_n[B^{\prime};\{1\}] - S_n[A^{\prime} \cup B^{\prime};\{1\}]
\nn\\
&& + \CO(\xe)
\eea
in the region $A^{\prime}_i:z <z_{2i-1}$ and the region $B^{\prime}_i:z>z_{2i}$
because all the local terms $s(a_i^{(n)})$ cancel each other.
\\

As discussed in the previous section, it remains unclear for what kind of states $\st{a^{(n)}}$
gives us finite and real boundary entropy, but the point is that as long as 
the states exist, we can prove that the mutual information is independent of the choice of centers.

\section{Conclusion}
\label{Sec:5}
\noindent 
We first studied the decomposition of Hilbert space in topological quantum field theory and the first-order formulation. These theories deserved a detailed mathematical analysis for the decomposition. Our analysis provided the supporting evidence for that in these theories one cannot obtain a direct sum decomposition or non-trivial centers in Hilbert space by removing operators in a trivial topology. The possible reason is that removing non-dynamical fields does not lead to a different decomposition. In the case of the first-order formulation, it is possibly due to the special form of quantization algebra. When we consider a non-trivial topology in topological quantum field theory, we also found it hard to keep centers only on an entangling surface by removing operators. Hence the mutual information possibly depends on the observational method. Our results also gave examples for that the constraints do not always give non-trivial centers after removing operators as both the topological quantum field theory and the first-order formulation have constraints in the quantization. 
\\

\noindent 
We also extensively studied the entanglement entropy in the non-interacting theory. Our computation provided supporting evidence that only in gauge theory, the universal term receives the contribution from the boundary entanglement entropy. This possibly points out that the presence of contact terms is due to the gauge symmetry. Hence our computation gave the motivation to further investigate the contact term in the gauge theory. We also compare two different computation methods. The first one is to regularize the zero modes of eigenfunctions in the heat kernel without imposing boundary conditions, and the second one is to consider classical boundary effects. Two methods should be equivalent for giving the same universal contribution to the entanglement entropy. The universal term of entanglement entropy in the $p$-form non-interacting theory in $2p+2$ dimensions is also expressed in terms of those of the massless non-interacting scalar field theory. We showed that the universal term of entanglement entropy is consistent with the anomaly coefficients \cite{CamporesiHiguchi} and also expect that this result should give us a better understanding of the holography. These results should help us learn more about the higher-dimensional conformal field theory. For the $p$-form non-interacting theory in $p+1$ dimensions, the universal term vanishes because these theories do not have any dynamical degrees of freedom. From the results of the $p$-form non-interacting theory in $p+1$ dimensions, the universal term of $p$-from non-interacting theory in $p+2$ dimensions is equivalent to that of the zero-form non-interacting theory in $p+2$-dimensions.\\

\noindent 
 In the non-interacting theory, the entanglement entropy is the sum of the classical Shannon entropy and bulk entanglement entropy \cite{Casini:2013rba}. Therefore, this provides some special properties that allowed us to prove the strong subadditivity \cite{Lieb:1973cp} generically. This result can possibly be extended to other cases like two-dimensional conformal field theory, in which the entanglement entropy has the same form as in the non-interacting theory. Therefore, it would be interesting to understand whether the conformal field or holographic gravity theory with the non-trivial center satisfies the strong subadditivity. 
\\

\noindent 
Finally, we computed the mutual information with centers. This computation sheds light on understanding whether mutual information depends on the choice of centers. Since the mutual information is in general hard to compute, we only considered two-dimensional conformal field theory. Our computation of two-dimensional conformal field theory was for the mutual information on a single interval and multiple intervals. The exact result offers supporting evidence that mutual information does not depend on the choice of centers in two-dimensional conformal field theory. 
\\

\noindent 
It is most interesting to consider the universal term of entanglement entropy, which does not depend on a regulator. There are many interesting questions for the universal term of quantum field theory that has not been studied. One question is whether the universal term of entanglement entropy depends on the choice of centers in the strong coupling limit. Because the holographic results provide a conjecture for that a strongly coupled conformal field theory is dual to a weakly coupled AdS gravity theory, the universal term follows from the minimal surface of bulk gravity theory. The minimum surface seems not to have a choice of center operators. Hence the universal term of entanglement entropy should not depend on the choice of centers in the strong coupling limit. The proof from a combination of the holography and operator languages was still unknown so far. 

\section*{Acknowledgment}
\noindent 
We would like to thank for Arpan Bhattacharyya, Horacio Casini, Dimitri Fursaev, Song He, Ling-Yan Hung, Charles Melby-Thompson, and Jie-Qiang Wu for their useful discussion.
\\

\noindent 
 Xing Huang was supported by the MOST Grant 103-2811-M-003-024 and the NSFC Grant No. 11947301. Chen-Te Ma was supported by the Post-Doctoral International Exchange Program and China Postdoctoral Science Foundation, Postdoctoral General Funding: Second Class (Grant No. 2019M652926). Chen-Te Ma would like to thank Nan-Peng Ma for his suggestion and encouragement. 
 \\
 
 \noindent 
We also would like to thank the Fudan University, Huazhong University of Science and Technology, Asia Pacific Center for Theoretical Physics at the Pohang University of Science and Technology, Yukawa Institute for Theoretical Physics at the Kyoto University, National Tsing Hua University, Tohoku University, Okinawa Institute of Science and Technology Graduate University, Istituto Nazionale Di Fisica Nucleare - Sezione di Napoli at the Università degli Studi di Napoli Federico II, Kadanoff Center for Theoretical Physics at the University of Chicago, Stanford Institute for Theoretical Physics at the Stanford University, Kavli Institute for Theoretical Physics at the University of California of the Santa Barbara, Israel Institute for Advanced Studies at the Hebrew University of
Jerusalem, Jinan University, Institute of Physics at the University of Amsterdam, Shing-Tung Yau Center at the Southeast University, Institute of Theoretical Physics at the Chinese Academy of Sciences, Shanghai University, Shanghai Jiao Tong University, Sun Yat-Sen University, Institute for Advanced Study at the Tsinghua University, Yangzhou University, and Zhejiang Institute of Modern Physics at the Zhejiang University.

\appendix
\section{Review of the von Neumann Algebra in\\
 the Entanglement with Centers}
\label{app1}
We review the entanglement from the algebraic point of view
\cite{Casini:2013rba, Ma:2015xes} and will particularly emphasize the role of von Neumann algebra, which is generically assumed in local quantum field theory. Now we consider the following operator algebras in two spatial regions ($V$ and its complement $\bar{V}$) satisfying:
\bea
A_V=A_{\bar{V}}^{\prime}, \qquad A_{\bar{V}}=A_V^{\prime},
\eea
where $A_V$ is algebra in the region $V$ and $A_{\bar{V}}$ is algebra in the region $\bar{V}$. We also denote the algebra $A^{\prime}$ as the commutant of the algebra $A$. The von Neumann algebra $A$ satisfies $A=A^{\prime\prime}$. We assume that the algebras $A_V, A_{\bar{V}}$ are von Neumann algebra.   
The tensor product decomposition of Hilbert space corresponds to the so-called
trivial center, in which case the only operator to commute with all operators is the identity operator.  
We should also include the non-trivial centers in our discussion without loss
of generality. Under the assumption of $A_V, A_{\bar{V}}$ being von Neumann algebra, the non-trivial
centers in a Hilbert space imply that the Hilbert space has no tensor product decomposition.
The local quantum field theory naturally has a trivial center. To construct non-trivial centers in local quantum field theory, we remove operators on an entangling surface. For example, we can remove momentum operators on an entangling surface, and then the position operators on the entangling surface become centers. We can interpret that choosing an entangling surface losing momentum operators is equivalent to not observing the momentum operators on the entangling surface. Removing operators or choosing centers can be viewed as the choice of measurements on an entangling surface (see \cite{Ma:2015xes} for more discussions). This gives a more general definition of entanglement.
\\

We first discuss how to define a reduced density matrix in the presence of non-trivial centers. First of all, we find a basis to diagonalize the center as the matrix
\begin{equation} 
Z\equiv\left(\begin{array}{cccc}
\lambda^11 & 0 & \hdots & 0 \\
0 & \lambda^21 & \hdots & 0 \\
\vdots & \vdots & &\vdots \\
0 & 0 & \hdots & \lambda^m 1
\end{array}\right)\,.
\end{equation}   
The algebra ($A\cup A^{\prime}$) is then isomorphic to the matrix
\bea
\left(\begin{array}{cccc}
A_1\otimes A_1^{\prime} & 0 & \hdots & 0 \\
0 & A_2\otimes A_2^{\prime} & \hdots & 0 \\
\vdots & \vdots & &\vdots \\
0 & 0 & \hdots & A_m\otimes A_m^{\prime}
\end{array}\right)\  ,
\eea
 and the algebra $A$ also takes a block-diagonal form as that
\bea
\left(\begin{array}{cccc}
A_1 & 0 & \hdots & 0 \\
0 & A_2 & \hdots & 0 \\
\vdots & \vdots & &\vdots \\
0 & 0 & \hdots & A_m
\end{array}\right)\,.
\eea
Therefore, the total Hilbert space $H$ is isomorphic to the Hilbert space $\bigoplus_k \bigg(H_V^k\otimes H_{\bar{V}}^k\bigg)$. Although the decomposition, in which we will call direct sum decomposition henceforth, is not a tensor product decomposition, it is possible to perform a partial trace operation in each subspace to define a reduced density matrix \cite{Casini:2013rba, Ma:2015xes}. 
\\

The reduced density matrix in the region $V$ is:
\bea
\mbox{Tr}_{\bar{V}}\rho_{A_VA_{\bar{V}}}=\rho_{A_V}=
\left(\begin{array}{cccc}
p_1\rho_{A_1} & 0 & \hdots & 0 \\
0 & p_2\rho_{A_2} & \hdots & 0 \\
\vdots & \vdots & &\vdots \\
0 & 0 & \hdots & p_m\rho_{A_m}
\end{array}\right)\,,
\nn\\
\eea
where $\mbox{Tr}\rho_{A_k}=1$, $p_m$ is the probability of center, and $\mbox{Tr}_{\bar{V}}$ denotes a partial trace operation over the region $\bar{V}$. The entanglement entropy $S_{EE}(A)\equiv -\mbox{Tr}\big(\rho_A\ln\rho_A\big)$ is then given by:
\bea
\label{eewithshannon}
-\mbox{Tr}\big(\rho_A\ln\rho_A\big)
=-\sum_k p_k\ln p_k-\sum_k\mbox{Tr}\big(p_k\rho_{A_k}\ln\rho_{A_k}\big).
\nn\\
\eea
The first term is the classical Shannon entropy, and the second term is the average entanglement entropy. If we consider centers with a continuous distribution, the classical Shannon entropy becomes
\bea
&&-\int_{\phi} \big(f(\phi)\Delta\big)\ln(f(\phi)\Delta)
\nn\\
&&\longrightarrow-\ln(\Delta)-\int d\phi\ f(\phi)\ln f(\phi),
\eea
in where we replace $p_k$ by $f(\phi)\Delta$ ($\xD$ for normalization $\int_\phi f(\phi)\xD = 1$). The classical Shannon entropy with the continuous distribution depends on $\Delta$ or the regularization schemes, and therefore the entanglement entropy can be negative. The second term in the classical Shannon entropy with the continuous distribution is called continuous entropy. We only consider the continuous entropy in the classical Shannon entropy with the continuous distribution. The mutual information $M(A, B)\equiv S_{EE}(A)+S_{EE}(B)-S_{EE}(A\cup B)$ is a suitable quantity for avoiding the regulators.

\section{Review of the Lagrangian Formulation in\\
 the Entanglement with Centers}
\label{app2}
We quickly go through the Lagrangian method for computing the entanglement entropy with centers and also review the replica trick and conical method.

\subsection{Lagrangian Method}
Defining entanglement entropy with non-trivial centers is the same as removing some operators to let remaining operators on an entangling surface commute with all operators in a Hilbert space. This acting of operator removing results in the suppression of quantum fluctuation on an entangling surface. Therefore, we adopt an on-shell action \cite{Ma:2015xes, Donnelly:2015hxa} and consider only quantum fluctuation in the bulk.
In the non-interacting theory, the bulk and boundary entanglement entropy can be separated, which can be seen from the Hamiltonian formulation. In the interacting theory, the boundary fields do not decouple from the bulk fields. Nevertheless, entanglement entropy in the interacting theory is also in the form of \er{eewithshannon} as the sum of classical Shannon entropy and average entanglement entropy, but the reduced density matrix $\rho_{A_k}$
generally depends on the value $k$ of centers. When one chooses a center of operators on an entangling surface, it is equivalent to choosing a boundary condition in the Lagrangian method \cite{Ma:2015xes, Donnelly:2015hxa}. We adopt this way to calculate entanglement entropy from the Lagrangian method.

\subsubsection{Replica Trick and Conical Method}
To compute the entanglement entropy, we use the replica trick or conical method in an $n$-sheet manifold. The entanglement entropy is rewritten as the following:
\bea
S_A=\lim_{n\rightarrow 1}\frac{\mbox{Tr}(\rho_A^n)-1}{1-n}=-\frac{\partial}{\partial n}\mbox{Tr}\rho_A^n\bigg|_{n=1}.
\eea
In order to compute the quantity $\mbox{Tr}\rho_A^n$, one considers the $n$ copies
\bea
(\rho_A)_{\phi_{1+}\phi_{1-}}(\rho_A)_{\phi_{2+}\phi_{2-}}\cdots (\rho_A)_{\phi_{n+}\phi_{n-}}
\eea
with the $n$-sheet boundary condition $\phi_{i-}=\phi_{(i+1)+}$. Then the path integral representation for the quantity $\mbox{Tr}\rho_A^n$ in the $n$-sheet manifold is given by 
\bea
\mbox{Tr}\rho_A^n=(Z_1)^{-n}\int D\phi\ e^{-S(\phi)},
\eea
where $(Z_1)^{-n}$ is inserted to normalize the reduced density matrix. The entanglement entropy from the conical method is given by
\bea
S_A=\bigg(1-\beta\frac{\partial}{\partial\beta}\bigg)\ln Z(\beta)\bigg|_{\beta=2\pi}\,.
\eea
Two methods are equivalent with the identification of $\beta=2\pi n$ $\big(Z(\beta)=Z(2\pi n)=Z(2\pi)^n\mbox{Tr}\rho_A^n\big)$, and these methods help us to define the entanglement entropy in an infinite dimensional Hilbert space.

\section{Review of Boundary Entanglement Entropy in\\ the Abelian One-Form Gauge Theory}
\label{app3}
The action for the Abelian one-form gauge theory in the Euclidean spacetime is given by
\bea
S_\zt{EAOG}=\frac{1}{4}\int d^Dx\ \sqrt{\det g_{\rho\sigma}}\ F_{\mu\nu}F^{\mu\nu},
\eea
where $g_{\mu\nu}$ is the metric field, $F_{\mu\nu}\equiv\nabla_{\mu}A_{\nu}-\nabla_{\nu}A_{\mu}$, and $\nabla_{\mu}$ is the covariant derivative of $\mu$ direction. When computing entanglement entropy with non-trivial centers, we need to choose a classical (in the sense
of no fluctuation) entangling surface. Here we do not include gauge fixing and ghost terms because these terms are not relevant for finding the on-sell boundary action. We split the one-form gauge field as that $A_{\mu}=A^{CL}_{\mu}+A^Q_{\mu}$, where $A^{CL}_{\mu}$ is a classical solution, which is compatible with a boundary condition, and $A^Q_{\mu}$ is the quantum fluctuation which vanishes on the entangling surface. The action is:
\bea
&&S_\zt{EAOG}
\nn\\
&=&\frac{1}{4}\int d^Dx\ \sqrt{\det g_{\rho\sigma}}\ F_{\mu\nu}F^{\mu\nu}
\nn\\
&=&\frac{1}{2}\int d^Dx\ \sqrt{\det g_{\rho\sigma}}\
\nn\\
&&\times \bigg(\nabla_{\mu}A^{CL}_{\nu}\nabla^{\mu}A^{CL}{}^{\nu}-\nabla_{\mu}A^{CL}_{\nu}\nabla^{\nu}A^{CL}{}^{\mu}
\nn\\
&&+\nabla_{\mu}A^{Q}_{\nu}\nabla^{\mu}A^{Q}{}^{\nu} -\nabla_{\mu}A^{Q}_{\nu}\nabla^{\nu}A^{Q}{}^{\mu}\bigg),
\eea
in which we used $\nabla_{\mu}g_{\nu\rho}=0$, $\nabla_{\mu}F^{CL\mu\nu}=0$, $F^{CL}_{\mu\nu}\equiv\nabla_{\mu}A^{CL}_{\nu}-\nabla_{\nu}A^{CL}_{\mu}$, $\hat{n}_{\nu^{\prime}}$, and $\nabla_{\mu} V^{\mu}=\bigg(1/\sqrt{\det g_{\nu\rho}}\bigg)\partial_{\mu}\bigg(\sqrt{\det g_{\sigma\delta}}V^{\mu}\bigg)$. The action can be rewritten with a boundary term as the following
\bea
&&\frac{1}{2}\int d^Dx\ \sqrt{\det g_{\rho\sigma}}\
\nn\\
&&\times \bigg(\nabla_{\mu}A^{CL}_{\nu}\nabla^{\mu}A^{CL}{}^{\nu}-\nabla_{\mu}A^{CL}_{\nu}\nabla^{\nu}A^{CL}{}^{\mu}
\nn\\
&&+\nabla_{\mu}A^{Q}_{\nu}\nabla^{\mu}A^{Q}{}^{\nu}-\nabla_{\mu}A^{Q}_{\nu}\nabla^{\nu}A^{Q}{}^{\mu}\bigg)
\nn\\
&=&\frac{1}{2}\int d^Dx\ \sqrt{\det g_{\rho\sigma}}\ 
\nn\\
&&\times\bigg(-A_{\nu}^{Q}\nabla_{\mu}\nabla^{\mu}A^{Q}{}^{\nu}+A^{Q}_{\nu}\nabla_{\mu}\nabla^{\nu}A^{Q}{}^{\mu}\bigg)
\nn\\
&&+\frac{1}{2}\int d^{D-1}x\ \hat{n}_{\mu^{\prime}}\sqrt{\det h_{\rho^{\prime}\sigma^{\prime}}}\ \bigg(A^{CL}_{\nu^{\prime}}F^{CL}{}^{\mu^{\prime}\nu^{\prime}}\bigg),
\eea
where $\hat{n}_{\nu^{\prime}}$ is the unit normal vector, and the induced metric $h_{\mu^{\prime}\nu^{\prime}}$ is defined by the following: 
\bea
ds^2&=&g_{\mu\nu}dx^{\mu}dx^{\nu}=g_{\mu\nu}\frac{\partial x^{\mu}}{\partial x^{\mu^{\prime}}}\frac{\partial x^{\nu}}{\partial x^{\nu^{\prime}}}dx^{\mu^{\prime}}dx^{\nu^{\prime}}
\nn\\
&\equiv& h_{\mu^{\prime}\nu^{\prime}}dx^{\mu^{\prime}}dx^{\nu^{\prime}}.
\eea
The boundary spacetime indices are labeled by the Greek indices with primes. Now we compute the boundary on-shell action. We first find an asymptotic solution. The metric is 
\bea
ds^2=dr^2+r^2d\theta^2+dx_{\bot}^2,
\eea
where the period of $\theta$ is $\beta$, and $x_{\bot}$ is the orthogonal coordinates. We choose the solution of the gauge field
\bea
A^{CL}_{\theta}=\sum_{n^{\prime}}\phi_{n^{\prime}}(r)\psi_{n^{\prime}}(x_{\bot}),
\eea
where
\bea
\int_{x_{\bot}}\psi_{m^{\prime}}\psi_{n^{\prime}}&=&\delta_{m^{\prime}n^{\prime}},
\nn\\
 \nabla^2\psi_{n^{\prime}}(x_{\bot})&=&-\lambda_{n^{\prime}}\psi_{n^{\prime}}(x_{\bot}).
\eea
The equation of motion is: 
\bea
\nabla_{\mu}F^{CL}{}^{\mu\nu}=\partial_{\mu}F^{CL}{}^{\mu\nu}+\Gamma^{\mu}{}_{\mu\rho}F^{CL}{}^{\rho\nu}=0,
\eea
 where
\bea
\Gamma^{\mu}{}_{\nu\rho}\equiv\frac{1}{2}g^{\mu\sigma}(\partial_{\rho}g_{\sigma\nu}+\partial_{\nu}g_{\sigma\rho}-\partial_{\sigma}g_{\nu\rho}).
\eea
We need to solve three equations. The first equation near the boundary is 
\bea
\partial_{\theta}F^{CL}{}^{\theta r}=0,
\eea 
the second equation near the boundary is 
\bea
\partial_{\theta}F^{CL}{}^{\theta x_{\bot}}+\partial_rF^{CL}{}^{rx_{\bot}}=0,
\eea
 and the final equation near the boundary is 
 \bea
 \partial_rF^{CL}{}^{r\theta}+\frac{1}{r}F^{CL}{}^{r\theta}+\partial_{x_{\bot}}F^{CL}{}^{x_{\bot}\theta}=0.
 \eea 
 Because the classical field strength does not depend on $\theta$ near the boundary, the first equation should be satisfied. Due to that the equation $F^{rx_{\bot}}=0$ satisfies the condition near the boundary, the second equation is also satisfied. Hence we only need to consider the last equation, which
gives:
 \bea
&&\sum_{n^{\prime}} \partial_r\bigg(\frac{1}{r^2}\partial_r\phi_{n^{\prime}}\bigg)\psi_{n^{\prime}}+ \frac{1}{r^3}\partial_r\phi_{n^{\prime}}\psi_{n^{\prime}}+\frac{1}{r^2}\phi_{n^{\prime}}\nabla^2\psi_{n^{\prime}}
\nn\\
&=&\sum_{n^{\prime}}\bigg(-\frac{1}{r^3}\partial_r\phi_{n^{\prime}}+\frac{1}{r^2}\partial_r^2\phi_{n^{\prime}}\bigg)\psi_{n^{\prime}}-\frac{1}{r^2}\phi_{n^{\prime}}\lambda_{n^{\prime}}\psi_{n^{\prime}}
\nn\\
&=&0.
\eea
Now we rewrite the equation above as the following
\bea
\frac{d^2}{d r^2}\phi_{n^{\prime}}-\frac{1}{r}\frac{d}{dr}\phi_{n^{\prime}}-\lambda_{n^{\prime}}\phi_{n^{\prime}}=0.
\eea
We then choose $\phi_{n^{\prime}}\approx a_{n^{\prime}}+b_{n^{\prime}} r^2\ln r$ near the boundary. 
Hence we obtain 
\bea
2b_{n^{\prime}}-\lambda_{n^{\prime}}a_{n^{\prime}}-\lambda_{n^{\prime}}b_{n^{\prime}}r^2\ln r=0
\eea
near the boundary. Because we set the boundary at $r=0$, we get the relation 
\bea
2b_{n^{\prime}}=\lambda_{n^{\prime}}a_{n^{\prime}}
\eea
 near the boundary. Now we define an electric field in terms of $n^{\prime}$ near boundary as the followings:
\bea
F^{CL}{}^{\theta r}\sim\frac{1}{r^2}F^{CL}_{\theta r}=-\frac{1}{r^2}\sum_{n^{\prime}}\partial_r\phi_{n^{\prime}}\psi_{n^{\prime}},
\eea
\bea
E_{n^{\prime}}&\equiv&-\frac{1}{r}\partial_r\phi_{n^{\prime}}\big|_{r=\epsilon\rightarrow 0}=-\frac{b_{n^{\prime}}}{r}(2r\ln r+r)\big|_{r=\epsilon\rightarrow 0}
\nn\\
&\sim& 2b_{n^{\prime}}\ln(\epsilon^{-1}),
\eea
\bea
E^{CB}\equiv F^{CL}{}^{\theta r}=\frac{1}{r}\sum_{n^{\prime}}E_{n^{\prime}}\psi_{n^{\prime}}.
\eea
We determine $\phi_{n^{\prime}}$ in terms of $E_{n^{\prime}}$ as the following:
\bea
\phi_{n^{\prime}}\big|_{r=\epsilon\rightarrow 0}=a_{n^{\prime}}=\frac{2b_{n^{\prime}}}{\lambda_{n^{\prime}}}\sim\frac{E_{n^{\prime}}}{\lambda_{n^{\prime}}\ln\epsilon^{-1}}.
\eea
Hence we compute the on-shell boundary action as the belows:
\bea
&&\frac{1}{2}\int d^{D-1}x\ \hat{n}_{\nu^{\prime}}\sqrt{\det h_{\rho^{\prime}\sigma^{\prime}}}\bigg(A^{CL}_{\nu^{\prime}}F^{CL}{}^{\mu^{\prime}\nu^{\prime}}\bigg)
\nn\\
&=&-\frac{1}{2}\int d\theta\sum_{n^{\prime}}\phi_{n^{\prime}}E_{n^{\prime}}
\nn\\
&\sim&-\sum_{n^{\prime}}\frac{\beta E_{n^{\prime}}^2}{2\lambda_{n^{\prime}}\ln\epsilon^{-1}}.
\eea
We note that that if $\lambda_{n^{\prime}}=0$, the on-shell action vanishes because $E_{n^{\prime}}=0$. 
Our measure is defined as the following:
\bea
\int DA_{\mu}\sim\int DA^Q_{\mu}DA^{CL}_{\nu}\sim \int DA^Q_{\mu}DE_{n^{\prime}}.
\eea
Because the boundary field does not couple to the bulk field, the partition function is the product of the partition function of  bulk field and that of boundary field. In the one-form Abelian gauge theory, the partition function of boundary field is given by:
\bea
Z^\zt{BOAG}&\sim&\Pi_{n^{\prime\prime}}\bigg(-\frac{\ln(\epsilon^{-1})\lambda_{n^{\prime\prime}}}{\beta}\bigg)^{\frac{1}{2}}
\nn\\
&=&\det{}^{\prime} \bigg(\frac{\ln(\epsilon^{-1})\nabla_{x_{\bot}}^2}{\beta}\bigg)^{\frac{1}{2}},
\eea
in which we exclude the modes with $\lambda_{n^{\prime\prime}}=0$, and the operation $\det{}^{\prime}$ only has the products of non-zero eigenvalues. Then we can use the conical method to obtain the boundary entanglement entropy in the Abelian one-form gauge theory.

\section{Bulk Entanglement Entropy in\\ the Abelian $p$-From Gauge Theory}
\label{app4}
We start by discussing the Abelian one-form gauge theory, and then we generalize the results to the Abelian $p$-form gauge theory. The action of the Abelian one-form gauge theory is given by
\bea
&&S_\zt{AOG}
\nn\\
&=&\int d^Dx\ \sqrt{\det g_{\rho\sigma}}\bigg(\frac{1}{4}F_{\mu\nu}F^{\mu\nu}+\frac{1}{2}\big(\nabla_{\mu}A^{\mu}\big)^2-\bar{c}\Box c\bigg),
\nn\\
\eea
in which we introduce the gauge fixing term and ghost field. Although the ghost field does not couple with the gauge field, the ghost field couples to the metric. Therefore, the ghost field should affect the results of entanglement entropy.
The bulk action is 
\bea
&&\int d^Dx\ \sqrt{\det g_{\rho\sigma}}\ 
\nn\\
&&\bigg(-\frac{1}{2}A^{\mu}\big(g_{\mu\nu}\Box+\lbrack\nabla_{\mu}, \nabla_{\nu}\rbrack\big)A^{\nu}-\bar{c}\Box c\bigg).
\eea
The free energy on the bulk is given by
\bea
\frac{1}{2}\ln\det\big(-g_{\mu\nu}\Box-\lbrack\nabla_{\mu}, \nabla_{\nu}\rbrack\big)-\ln\det(-\Box).
\eea
To compute the free energy, we introduce the complete basis $\phi_n$ as that
\bea
-\Box\phi_n=\lambda_n\phi_n.
\eea
Then we could define a scalar heat kernel as the following
\bea
K_s(s, x, y)=\sum_ne^{-s\lambda_n}\phi_n(x)\phi_n(y).
\eea
We also introduce a complete set of eigenfunctions as the below
\bea
\big(-g_{\mu\nu}\Box-\lbrack\nabla_{\mu}, \nabla_{\nu}\rbrack\big)A_n^{\nu}=\lambda_nA_{n, \mu}.
\eea
Then the one-form gauge heat kernel is defined as the following
\bea
K_{og}(s, x, y)_{\mu\nu}=\sum_ne^{-s\lambda_n}A_{\mu, n}(x)A_{\nu, n}(y).
\eea
The scalar eigenfunctions could be used to express the one-form gauge eigenfunctions as the following:
\bea
A^L_{\mu^{\prime}, n}&=&\frac{1}{\sqrt{\lambda^{\prime}_n}}\nabla_{\mu^{\prime}}\phi_n,\qquad A^T_{\mu^{\prime}, n}=\frac{1}{\sqrt{\lambda^{\prime}_n}}\epsilon_{\mu^{\prime}\nu^{\prime}}\nabla^{\nu^{\prime}}\phi_n, 
\nn\\
 A_{\mu^{\prime}, n}&=&A^L_{\mu^{\prime}, n}+A^T_{\mu^{\prime}, n},
\eea
and the other components of one-form gauge eigenfunctions are identified as the scalar eigenfunctions directly. The directions of two-dimensional cone are labeled by $\mu^{\prime}$. We define $\lambda_n^{\prime}$ as the below
\bea
-\nabla^{\mu^{\prime}}\nabla_{\mu^{\prime}}\phi_n=\lambda_n^{\prime}\phi_n.
\eea
Therefore, we obtain the followings:
\bea
&&K_{og}( s, x, x)_{\mu}^{\mu}
\nn\\
&=&\sum_n\bigg(\frac{e^{-s\lambda_n}}{\lambda^{\prime}_n}\big(2\nabla_{\mu^{\prime}}\phi_n\nabla^{\mu^{\prime}}\phi_n\big)\bigg)
\nn\\
&&+\sum_{n,\ \lambda_n\neq\lambda_n^{\prime}}(D-2)e^{-s\lambda_n}\phi_n^2
\nn\\
&=&\sum_n\bigg\lbrack\frac{e^{-s\lambda_n}}{\lambda^{\prime}_n}\bigg(-2\phi_n\nabla^{\mu^{\prime}}\nabla_{\mu^{\prime}}\phi_n
+2\nabla_{\mu^{\prime}}\big(\phi_n\nabla^{\mu^{\prime}}\phi_n\big)\bigg)\bigg\rbrack
\nn\\
&&+\sum_{n,\ \lambda_n\neq\lambda_n^{\prime}}(D-2)e^{-s\lambda_n}\phi_n^2
\nn\\
&=&\sum_n e^{-s\lambda_n}\bigg(2\phi_n^2+\frac{\nabla^{\mu^{\prime}}\nabla_{\mu^{\prime}}(\phi_n^2)}{\lambda^{\prime}_n}\bigg)
\nn\\
&&+\sum_{n,\ \lambda_n\neq\lambda_n^{\prime}}(D-2)e^{-s\lambda_n}\phi_n^2
\nn\\
&=&2K_s(s, x, x)+(D-2)K_s^{\prime}(s, x, x)
\nn\\
&&+\sum_n \frac{e^{-s\lambda_n}}{\lambda^{\prime}_n}\nabla^{\mu^{\prime}}\nabla_{\mu^{\prime}}(\phi_n^2),
\eea
where $K_s^{\prime}$ is the heat kernel that does not include the zero modes of transverse directions (a unit sphere). When we consider the four dimensional one-form Abelian gauge theory in the Euclidean flat background with $S^2$ entangling surface, the zero mode of transverse directions should not have any contribution to the universal term of entanglement entropy. Because $\nabla\cdot\vec{E}=\nabla\cdot\vec{B}=0$, where $\vec{E}$ is the electric field, and $\vec{B}$ is the magnetic field, we have the following term:
\bea
\frac{\partial E_{l=0}}{\partial r}+\frac{2}{r}E_{l=0}=\frac{\partial B_{l=0}}{\partial r}+\frac{2}{r}B_{l=0}=0,
\eea
where $E_{l=0}$ is the electric field for the zero mode in the transverse directions, and $B_{l=0}$ is the magnetic field for the zero mode of the transverse directions. As a result, we need the zero electric and magnetic fields in order to have finite energy. This implies that the zero mode in the transverse directions does not give universal contributions to the entanglement entropy. In other words, we cannot naively use the scalar field to replace the one-form Abelian gauge field due to the over-counting of zero mode in the transverse directions \cite{Casini:2015dsg}.
\\

The free energy is given by
\bea
&&-\frac{1}{2}\int_{\epsilon^2}^{\infty}ds\ \frac{e^{-m^{\prime}{}^2s}}{s}\int d^Dx\ \sqrt{\det g_{\nu\rho}}\ 
\nn\\
&&\times\bigg(K_{og}(s, x, x)_{\mu}^{\mu}-2K_s(s, x, x)\bigg)
\nn\\
&=&-\frac{1}{2}\int_{\epsilon^2}^{\infty}ds\ \frac{e^{-m^{\prime}{}^2s}}{s}\int d^Dx\ \sqrt{\det g_{\mu\nu}}\
\nn\\
&&\times \bigg((D-2)K^{\prime}_s(s, x, x)+\sum_n \frac{e^{-s\lambda_n}}{\lambda^{\prime}_n}\nabla^{\mu^{\prime}}\nabla_{\mu^{\prime}}(\phi_n^2)\bigg),
\nn\\
\eea
where $m^{\prime}$ is an infrared regulator. The second term in the last equality is dominant for the zero modes of two-dimensional cone directions, and this term is also a total derivative term. Hence the second term in the last equality should correspond to the boundary entanglement entropy.
\\

Now we extend the computation of entanglement entropy to the $p$-form non-interacting theory. We first discuss how to introduce the ghost fields \cite{Obukhov:1982dt, Copeland:1984qk} for
computing the partition function of bulk fields in the $p$-form non-interacting theory. The action of the $p$-form non-interacting theory is
\bea
\frac{1}{2(p+1)!}(H, H),
\eea
where $H=dB$, $H$ is the field strength associated with the $p$-form field $B$, and the inner product between two $p$-forms in $D$ dimensions is defined as
\bea
(\alpha, \beta)\equiv\int d^Dx\ \big[\det(g_{\mu\nu})\big]^{\frac{1}{2}}\alpha^{\mu_1\mu_2\cdots\mu_p}\beta_{\mu_1\mu_2\cdots\mu_p}.
\eea
We can define the adjoint operator $\delta$ for the differential operator $d$ by using the inner product (when neglecting the boundary term)
\bea
(\alpha, d\beta)\equiv -(\delta\alpha, \beta),
\eea
where
\bea
\delta\alpha=\nabla^{\mu_1}\alpha_{\mu_1\mu_2\cdots\mu_p}dx^{\mu_2}\wedge dx^{\mu_2}\wedge\cdots dx^{\mu_p}\,.
\eea
The operators $\delta$ and $d$ also satisfy the following:
 \bea
 d^2=\delta^2=0.
 \eea
 The generalization of the Laplacian in the $p$-form non-interacting theory is defined as
 \bea
 \Box_p\equiv d\delta+\delta d.
 \eea
 Therefore, the action in the $p$-form non-interacting theory without considering a boundary term could be rewritten as the followings:
 \bea
 &&\frac{1}{2(p+1)!}(dB, dB)
 \nn\\
 &=&-\frac{1}{2(p+1)!}(B, \delta dB)
 \nn\\
 &=&-\frac{1}{2(p+1)!}\big(B, (\delta d+d\delta)B\big)
+\frac{1}{2(p+1)!}(B, d\delta B),
 \nn\\
 &=&-\frac{1}{2(p+1)!}(B, \Box_p B)
-\frac{1}{2(p+1)!}(\delta B,\delta B).
 \eea
 Then we introduce a gauge fixing term as that
 \bea
 \frac{1}{2(p+1)!}(\delta B, \delta B)
 \eea
 and a ghost action as that:
 \bea
 &&\frac{1}{2p!}(dG, \Box_p d G)
 \nn\\
 &=&\frac{1}{2p!}(dG, d\delta dG)
 \nn\\
 &=&-\frac{1}{2p!}(G, \delta d\delta d G)\nn\\
 & = & -\frac{1}{2 p!}\big(G, (\delta d\delta d+d\delta d\delta)G\big)+\frac{1}{2p!}(G, d\delta d\delta G)
 \nn\\
 &=&-\frac{1}{2p!}(G, \Box^2_{p-1} G)-\frac{1}{2p!}(\delta G, \delta d\delta G)\nn\\
 & = &- \frac{1}{2p!}(G, \Box_{p-1}^2 G)-\frac{1}{2p!}(\delta G, \Box_{p-2}\delta G).
 \eea
 Since the $(p-1)$-form ghost field also has its own gauge symmetry, we need to choose an additional gauge fixing term
 \bea
 \frac{1}{2p!}(\delta G, \Box_{p-2}\delta G)
 \eea
 and a ghost action 
 \bea
 \frac{1}{2(p-1)!}(d\tilde{G}, \Box_{p-1}^2d\tilde{G}),
 \eea 
 where $\tilde{G}$ is a commuting field to remove the non-physical degrees of freedom. It is necessary to continue the procedure until we encounter a zero-form ghost field, which does not have a gauge symmetry. We would also like to remind the reader that the ghost fields are commuting when they are ($p$-2$j$)-form fields and anti-commuting in other cases. In summary, the action of the $p$-form non-interacting theory is
 \bea
 S_p=-\frac{1}{2}\sum_{j=0}^p\frac{1}{(p+1-j)!}(A_{p-j}, \Box_{p-j}^{j+1} A_{p-j}),
 \eea
in which the $p$-form and ghost fields are denoted as $A_p$, and $A_{p-j}$ is commuting when $j$ is an even non-negative integer while anti-commuting for odd $j$. Hence the bulk partition function of the $p$-form non-interacting theory is determined as the following
\bea
\prod\limits_{j=0}^{p}\det\bigg(-\Box_{p-j}^{\big(\frac{j+1}{2}\big){(-1)^{{}^{(j+1)}}}}\bigg),
\eea
and we use the heat kernels to rewrite the bulk partition function as the following
\bea
&&-\frac{1}{2}\int_{\epsilon^2}^{\infty}ds\ \frac{e^{-m^{\prime 2}s}}{s}\int d^Dx\ \sqrt{\det g_{\mu\nu}}\ 
\nn\\
&&\times\mbox{Tr}\bigg(K_p(s, x, x)-2K_{p-1}(s, x, x)+\cdots 
\nn\\
&&+(-1)^p(p+1)K_0\bigg),
\eea
 where $K_p$ is defined as that
 \bea
 K_{p-j}(s, x, x^{\prime})=\bigg<x\bigg| e^{-s\bigg(-\Box_{(p-j)}\bigg)}\bigg|x^{\prime}\bigg>.
\eea
\\
 
The computation of entanglement entropy with a boundary term is equivalent to the computation of entanglement entropy considering the regularization of zero modes of the two-dimensional cone directions \cite{Kabat:1995eq}.

\section{Sphere and Two-Dimensional Cone}
\label{app5}
We first show that the computation of the entanglement entropy in the Euclidean polar coordinate with the entangling surface $S^{D-2}$ is equivalent to the case of $S^D$ with a unit radius in conformal field theory \cite{Casini:2011kv}. The Euclidean polar coordinate is
\bea
ds^2=dt^2+dr^2+r^2d\Omega^2_{D-2},
\eea
where $\Omega_{D-2}$ is the solid angle, and the entangling surface is at $r=R$ and $t=0$. We use a coordinate transformation:
\bea
t&=&R\frac{\sin\big(\frac{\tau}{R}\big)}{\cosh\big(u\big)+\cos\big(\frac{\tau}{R}\big)}, \qquad r=R\frac{\sinh\big(u\big)}{\cosh\big( u\big)+\cos\big(\frac{\tau}{R}\big)}, 
\nn\\
0&\le& u <\infty, \qquad 0\le\frac{\tau}{R}< 2\pi n
\eea
on the $n$-sheet manifold, which gives the following terms:
\bea
&&dt
\nn\\
&=&\frac{\bigg(1+\cos\big(\frac{\tau}{R}\big)\cosh\big(u\big)\bigg)d\tau}{\bigg(\cosh\big(u)+\cos\big(\frac{\tau}{R}\big)\bigg)^2}
\nn\\
&&-\frac{R\sin\big(\frac{\tau}{R}\big)\sinh\big(u\big)du}{\bigg(\cosh\big( u\big)+\cos\big(\frac{\tau}{R}\big)\bigg)^2},
\nn\\
&&d\rho
\nn\\
&=&\frac{R\bigg(1+\cos\big(\frac{\tau}{R}\big)\cosh\big(u\big)\bigg)du}{\bigg(\cosh\big(u)+\cos\big(\frac{\tau}{R}\big)\bigg)^2}
\nn\\
&&+\frac{\sin\big(\frac{\tau}{R}\big)\sinh\big(u\big)d\tau}{\bigg(\cosh\big( u\big)+\cos\big(\frac{\tau}{R}\big)\bigg)^2},
\nn\\
&&dt^2+d\rho^2
\nn\\
&=&\frac{d\tau^2}{\bigg(\cosh^2\big(u\big)+\cos^2\big(\frac{\tau}{R}\big)\bigg)^2}
\nn\\
&&+R^2\frac{du^2}{\bigg(\cosh^2\big(u\big)+\cos^2\big(\frac{\tau}{R}\big)\bigg)^2},
\eea
and then we obtain the metric in the new coordinate as that
\bea
&&ds^2
\nn\\
&=&\frac{d\tau^2}{\bigg(\cosh^2\big(u\big)+\cos^2\big(\frac{\tau}{R}\big)\bigg)^2}
\nn\\
&&+R^2\frac{du^2}{\bigg(\cosh^2\big(u\big)+\cos^2\big(\frac{\tau}{R}\big)\bigg)^2}
\nn\\
&&+R^2\frac{\sinh^2\big(u\big)d\Omega_{D-2}^2}{\bigg(\cosh^2\big(u\big)+\cos^2\big(\frac{\tau}{R}\big)\bigg)^2}.
\eea
In conformal field theory, we omit the common pre-factor, and the new metric becomes
\bea
ds_1^2 = \frac{d\tau^2}{R^2}+du^2+\sinh^2\big(u\big)d\Omega^2_{D-2}.
\eea 
Then we redefine  $\sinh\big(u\big)=\tan\big(\theta\big)$, where $0\le\theta <\pi/2$, and get
\bea
ds_2^2 = \frac{d\tau^2}{R^2}+\frac{d\theta^2}{\cos^2\big(\theta\big)}+\tan^2\big(\theta\big)d\Omega^2_{D-2},
\eea
in which we used
\bea
du^2=\frac{d\theta^2}{\cos^2\big(\theta\big)}.
\eea
In conformal field theory, we could omit the pre-factor and get
\bea
ds_3^2 = d\theta^2+\cos^2\big(\theta\big)\frac{d\tau^2}{R^2}+\sin^2\big(\theta\big)d\Omega^2_{D-2}.
\eea 
Then we could relate the sphere to a product geometry of a two-dimensional cone and a unit sphere near the entangling surface ($\theta\rightarrow\pi/2$) from a conformal mapping as the followings:
\bea
ds_4^2&=&\frac{1}{\sin^2(\theta)}d\theta^2+\frac{\cos^2(\theta)}{\sin^2(\theta)}\frac{d\tau^2}{R^2}+d\Omega^2_{D-2},
\nn\\
r&=&\frac{\cos(\theta)}{\sin(\theta)}, 
\nn\\
ds_4^2&=&\sin^2(\theta)dr^2+r^2\frac{d\tau^2}{R^2}+d\Omega^2_{D-2}, 
\nn\\
ds_4^2&\rightarrow&dr^2+r^2\frac{d\tau^2}{R^2}+d\Omega^2_{D-2}, \qquad \theta\rightarrow\frac{\pi}{2}.
\eea
This implies that we can introduce a two dimensional cone and set boundary condition in the cone to compute the boundary entanglement entropy on a sphere. Hence we identify a universal term from the partition function on a sphere and that of the entanglement entropy across a spherical entangling surface in the Euclidean flat space. When we use a sphere or a regularized cone to compute entanglement entropy, there is no a boundary term, but the effects of the boundary term also appear from the regularization.
\\

\end{document}